\documentclass{article}
\usepackage{ijcai23}

\usepackage{times}
\usepackage{soul}
\usepackage{url}
\usepackage[hidelinks]{hyperref}
\usepackage[utf8]{inputenc}
\usepackage[small]{caption}
\usepackage{graphicx}
\usepackage{amsmath}
\usepackage{amsthm}
\usepackage{booktabs}
\usepackage{algorithm}
\usepackage{algorithmic}
\usepackage[switch]{lineno}
\usepackage{tikz}
\usetikzlibrary{trees}
\usepackage{enumitem}

\usepackage{multirow}
\usepackage{multicol}

\newcommand{\stitle}[1]{\vspace{1ex} \noindent{\bf #1}}

\title{Poisoning Attacks against Recommender Systems: A Survey}

\author{
Zongwei Wang$^1$\and
Min Gao$^1$\footnote{Corresponding Author}\and
Junliang Yu$^2$\and
Hao Ma$^1$\and
Hongzhi Yin$^2$\and
Shazia Sadiq$^2$
\affiliations
$^1$Chongqing University \and
$^2$The University of Queensland
\emails
$^1$\{zongwei, gaomin, mahao\}@cqu.edu.cn \and
$^2$\{jl.yu@, h.yin1@, shazia@eecs.\}uq.edu.au
}

\begin{document}

\maketitle

\begin{abstract}
Modern recommender systems have seen substantial success, yet they remain vulnerable to malicious activities, notably poisoning attacks. These attacks involve injecting malicious data into the training datasets of RS, thereby compromising their integrity and manipulating recommendation outcomes for unscrupulous gains. This survey paper provides a systematic and up-to-date review of the research landscape on Poisoning Attacks against Recommendation (PAR). A novel and comprehensive taxonomy is proposed, categorizing existing PAR methodologies into three distinct categories: Component-Specific, Goal-Driven, and Capability Probing. Each category is examined in detail, covering their mechanisms and associated methods. To spur future research, we also highlight present challenges and propose innovative directions in this field. Moreover, to facilitate and standardize the empirical comparison of PAR, we introduce an open-source library, ARLib, which encompasses a comprehensive collection of PAR models and datasets. The library is released at \url{https://github.com/CoderWZW/ARLib}.
\end{abstract}

\section{Introduction}
Modern recommender systems (RS) \cite{42wang2023efficient,43cheng2016wide} capitalize the power of advanced deep neural architectures to help users discover items that align with their unique preferences. Despite their unprecedented effectiveness, there is a growing body of evidence pointing to the vulnerability of these deep RS that they are susceptible to malicious activities commonly referred to as \underline{P}oisoning \underline{A}ttacks against \underline{R}ecommendation (PAR)~\cite{50gunes2014shilling,51si2020shilling,85zhang2020gcn}. By introducing malicious or misleading data into the training data of RS, these attacks can disrupt the systems and manipulate the generated recommendations. For instance, training with poisoned data, the E-commerce RS would provide low-quality products with more exposure. Similarly, in the context of news dissemination, misinformation might be deliberately recommended to certain user groups. These PAR subvert the original purpose of RS, transforming them from efficient intermediaries between producers and consumers into tools manipulated for attackers' benefits. This scenario raises significant concerns regarding the security and integrity of RS. 

To safeguard RS against threats from PAR, it is imperative for researchers to understand the tactics used by potential attackers to achieve their objectives. Early investigations into PAR predominantly center on heuristic-based methods \cite{47o2002promoting,48yu2017hybrid,87burke2005limited}. These methods involved predefining a fixed strategy for generating and injecting malicious profiles into the system, crucial in evaluating the robustness of RS at the time. Nevertheless, such fixed construction patterns lacked adaptability, rendering them detectable to defense measures once their patterns are deciphered ~\cite{16wang2022gray}. 
\begin{figure}[]
\centering
\includegraphics[width=0.49\textwidth]{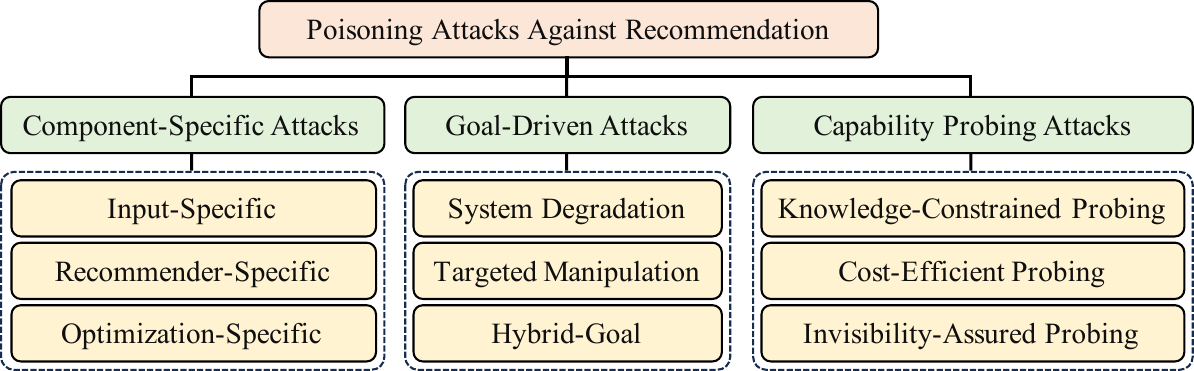}
\caption{The taxonomy of poisoning attacks against RS.}
\vspace{-15pt}
\label{Taxonomy}
\end{figure} 
Thus, there has been a paradigm shift towards more sophisticated and adaptable PAR. The latest trajectory of research on the PAR is \textbf{intrinsically aligned with the iterative advancements in recommendation techniques}. The evolution from factorization-based ~\cite{76koren2009matrix} to network-based ~\cite{02fang2018poisoning}, and currently to the cutting-edge large language model (LLM)-based recommendations ~\cite{77wu2023survey}, not only yields more precise recommendation outcomes but also introduces a range of new vulnerabilities. This compels researchers to identify novel PAR for varying RS contexts. Additionally, the study of PAR also involves \textbf{examining the objectives behind these attacks}, such as degradation of overall recommendation performance and influence on specific items or users. Given the varying attack measures and forms associated with different intended goals, recognizing and understanding these variances is vital for crafting effective defensive measures against such attacks. Lastly, \textbf{understanding the capabilities of attackers} is significant. In real-world scenarios, attackers often struggle to obtain detailed information about RS. Resource constraints further impede their ability to launch effective attacks while remaining undetected ~\cite{07lin2020attacking,74zeng2023practical}. Gaining insights into the challenges and limitations faced by attackers can stimulate the development of more effective defense strategies, ultimately enhancing the security of RS.

While previous surveys have specifically investigated PAR, they predominantly focus on the methodologies of constructing malicious profiles. Thus they lack a comprehensive perspective and fail to consider some crucial aspects of PAR, such as the diversity of attack objectives and the limitations of attacker capabilities. Given the rapid evolution of PAR and increasing concerns, there is an urgent need for an up-to-date and systematic survey encapsulating the latest advancements. In response, this paper presents a current and comprehensive landscape of PAR, discussing three distinct categories of PAR that involve attack strategies, objectives and capability. The corresponding taxonomy is illustrated in Figure~\ref{Taxonomy}. Additionally, to aid the implementation of research in this area, we develop an open-source \underline{Lib}rary for \underline{A}ttacks against \underline{R}ecommendation (ARLib). It integrates 15 influential PAR methods (with additional methods being continuously added), enabling the rapid development and comparison of PAR models. For details on implemented methods, please refer to the ARLib link. In summary, our contributions are threefold:
\begin{itemize}[leftmargin=*]
    \item Our survey on PAR introduces a unique and comprehensive taxonomy, which will deepen and broaden the understanding of PAR, offering a fresh perspective on the field.    
    \item Our survey encapsulates a large number of related and latest papers and sheds light on the current challenges and promising future directions. These critical insights will spur further innovation and exploration in PAR studies.
    \item We release an open-source library to facilitate the implementation and comparison of PAR methods, which is a valuable resource for researchers in this field. 
\end{itemize}

\stitle{Paper Collection.} In this survey, we conducted a thorough review of 45 latest high-quality research papers focusing on PAR. Our literature search was primarily conducted using DBLP and Google Scholar, with the keywords ``attack + recommendation" and ``poisoning + recommendation". We then traversed the citations of the identified papers and included relevant studies. To capture the most influential and cutting-edge work, we consistently monitored leading conferences and journals, including IJCAI, CIKM, KDD, WWW, SIGIR, WSDM, TKDE, etc. Additionally, to encompass emerging trends and innovative concepts, we included preprints from arXiv that presented novel ideas to this field.

\stitle{Connections to Existing Surveys.} To the best of our knowledge, while there are surveys investigating trustworthy ~\cite{52fan2022comprehensive} or robust  ~\cite{53anelli2021adversarial} RS, they primarily orient their focus towards defense mechanisms, leading to a deficit in specialization in the attack strategies. In addition, there are studies that have summarized heuristic-based PAR ~\cite{51si2020shilling} and those rooted in adversarial learning ~\cite{54deldjoo2021survey}. However, as research progresses, these perspectives fall short of encapsulating the current panorama of the PAR research. 


\section{Taxonomy of Poisoning Attacks against Recommendation}
In this section, the preliminaries of PAR are provided first, followed by the introduction of a novel and comprehensive taxonomy as illustrated in Figure~\ref{Taxonomy}.
\subsection{Preliminaries}
\stitle{Victim Recommender System.} Let $\mathcal{U} $ and $\mathcal{I} $ respectively be the sets of users and items in RS, and $\mathcal{D}$ denote the original user/item data, encompassing interaction information (e.g., clicks, likes, and collections) and features (e.g., text and images). The victim model in PAR refers to a recommender $f$ that aims to learn a low-dimensional representation set for predicting the preference of user $u \in \mathcal{U} $ to item $i \in \mathcal{I}$. We formulate the victim (recommendation) task as:

\begin{equation}
\begin{aligned}
\mathbf{\Theta}^{*}=\arg\min\limits_{\mathbf{\Theta}}\mathcal{L}_{rec}(f(\mathcal{D})),
\end{aligned}
\label{bi-level optimization}
\end{equation}
where $\mathbf{\Theta}^{*}$ represents the optimal model parameters, including user and item representations, and $\mathcal{L}_{rec}$ is the recommendation objective function such as Cross-Entropy.

\stitle{Poisoning Attacks against Recommendation.} In the setting of the PAR, attackers can manipulate a set of malicious users $\mathcal{U}_{M}$. These users engage in strategic interactions with various items, occasionally involving modifications to the features of some items. The resultant malicious data $\mathcal{D}_{M}$ is then injected into the training data of RS. The intent behind such attacks is to manipulate recommendation models' learning process and ultimately achieve specific objectives. Given this context, we formulate the PAR task as a bi-level optimization problem:

\begin{equation}
\begin{aligned}
&\mathcal{D}_{M}=\arg\max\limits_{\mathcal{D}_{M}}\mathcal{L}_{attack}(\mathcal{D}, \mathcal{D}_{M}, \mathbf{\Theta}^{*}), \\
\mathrm{s.t.}&, \quad  \mathbf{\Theta}^{*}=\arg\min\limits_{\mathbf{\Theta}}\mathcal{L}_{rec}(f(\mathcal{D},\mathcal{D}_{M})),
\end{aligned}
\label{bi-level optimization}
\end{equation}
where $\mathcal{L}_{attack}$ is the poisoning attack loss for evaluating attack utility, typically including measures such as the decrease in recommendation accuracy and the disruption in the ranking of specific target items. The poisonous data $\mathcal{D}_{M}$ are generated by alternately performing the inner optimizations for victim model training and outer optimizations. Specifically, the inner optimization focuses on training the victim model by considering both the original data and poisonous data. Meanwhile, the outer optimization is tasked with refining the poisonous data to achieve the intended attack objectives. Note that $f$ here can be a surrogate model as attackers might have no access to the victim model.

\subsection{Proposed Taxonomy}
In the domain of PAR, we categorize poisoning attacks into three distinct groups based on their patterns and intents, shown in Figure \ref{Taxonomy}. We also underline the symbols which play crucial roles in determining the characteristics of these three categories in their formulations shown in Equation (\ref{csattack-eq}-\ref{cpattack-eq}).

\stitle{Component-Specific Attacks.} Due to the rapid evolution of deep learning (DL) techniques, the landscape of RS now encompasses a diverse array of input data types (graphs, text, image, etc), recommender architectures (factorization-based, network-based, language model-based, etc), as well as optimization functions (Cross-Entropy, Softmax, Triplet loss, etc). The distinct characteristics of these essential components potentially give rise to varied forms of poisoning attacks. Recognizing the critical importance of this issue, a growing number of researchers are dedicating to exploring and identifying vulnerabilities specific to different components in RS, including input data, recommender architectures, and optimization processes. We formulate the component-specific PAR as:
\begin{equation}
\begin{aligned}
&\underline{\tilde{\mathcal{D}}_{M}}=\arg\max\limits_{\underline{\tilde{\mathcal{D}}_{M}}}\mathcal{L}_{attack}(\underline{\tilde{\mathcal{D}}}, \underline{\tilde{\mathcal{D}}_{M}}, \mathbf{\Theta}^{*}), \\
\mathrm{s.t.}&, \quad  \mathbf{\Theta}^{*}=\arg\min\limits_{\mathbf{\Theta}}\tilde{\mathcal{L}}_{rec}(\underline{\tilde{f}}(\underline{\tilde{\mathcal{D}}},\underline{\tilde{\mathcal{D}}_{M}})),
\end{aligned}
\label{csattack-eq}
\end{equation}
where $\tilde{\mathcal{D}}$ and $\tilde{\mathcal{D}}_{M}$ denotes the dataset comprising data of a specific type, $\tilde{f}$ represents the victim model with a particular recommender architecture and $\tilde{\mathcal{L}}_{rec}$ refers to a specific objective function.
        
\stitle{Goal-Driven Attacks.} This category is established based on the varying objectives that attackers pursue in their attack strategies. For instance, some attackers, fueled by the inter-platform competition, might be motivated to impair the overall system performance. Their primary aim is to degrade the user experience, creating a negative perception of the platform's efficiency and reliability. On another front, attackers driven by personal or business interests might strategically manipulate recommendation algorithms to artificially promote or demote the visibility and ranking of specific items. This could include various forms of content such as news articles, videos, or product listings, with the intention of broadcasting certain information to a wide user base or targeting specific user groups. Furthermore, sometimes these two types of objectives might be coupled to form a hybrid attack strategy in order to gain more benefits. Given that varying objectives impart distinct characteristics to PAR, it is also significant to examine PAR from the goal-driven perspective. we formulate this type of attacks as:
\begin{equation}
\begin{aligned}
&\mathcal{D}_{M}=\arg\max\limits_{\mathcal{D}_{M}}\underline{\tilde{\mathcal{L}}_{attack}}(\mathcal{D}, \mathcal{D}_{M}, \mathbf{\Theta}^{*}), \\
\mathrm{s.t.}&, \quad  \mathbf{\Theta}^{*}=\arg\min\limits_{\mathbf{\Theta}}\mathcal{L}_{rec}(f(\mathcal{D},\mathcal{D}_{M})),
\end{aligned}
\label{gdattack-eq}
\end{equation}
where $\tilde{\mathcal{L}}_{attack}$ denotes the loss function that assesses the utility of the attack in relation to specific attack goals.
        
\stitle{Capability Probing Attacks.} In practical scenarios, attackers are often constrained by various limitations. Commonly, they lack full access to the data and model details of RS. Meanwhile, financial or resource costs also prevent them from injecting excessive users and interactions. Additionally, attackers must remain vigilant against detection measures, which adds complexity to their attack strategies. Understanding these limitations faced by attackers is crucial for building robust RS and consolidating defense measures. Therefore, researchers are actively investigating the effect of PAR under varying levels of attacker capabilities. We define this category of PAR as:
\begin{equation}
\begin{aligned}
\underline{\mathcal{D}^{Limited}_{M}}&=\arg\max\limits_{\mathcal{D}^{Limited}_{M}}\mathcal{L}_{attack}(\underline{\mathcal{K}}(\mathcal{D}, \underline{\mathcal{D}^{Limited}_{M}}, \mathbf{\Theta}^{*})), \\
\mathrm{s.t.}, \quad  &\mathbf{\Theta}^{*}=\arg\min\limits_{\mathbf{\Theta}}\mathcal{L}_{rec}(f(\mathcal{D},\underline{\mathcal{D}^{Limited}_{M}})), \\
\end{aligned}
\label{cpattack-eq}
\end{equation}
where $\mathcal{D}^{Limited}_{M}$ is the malicious dataset that complies with specific limitations, and $\mathcal{K}$ represents the constraints related to knowledge of data and model details.

\begin{figure}[t]
\centering
\includegraphics[width=0.46\textwidth]{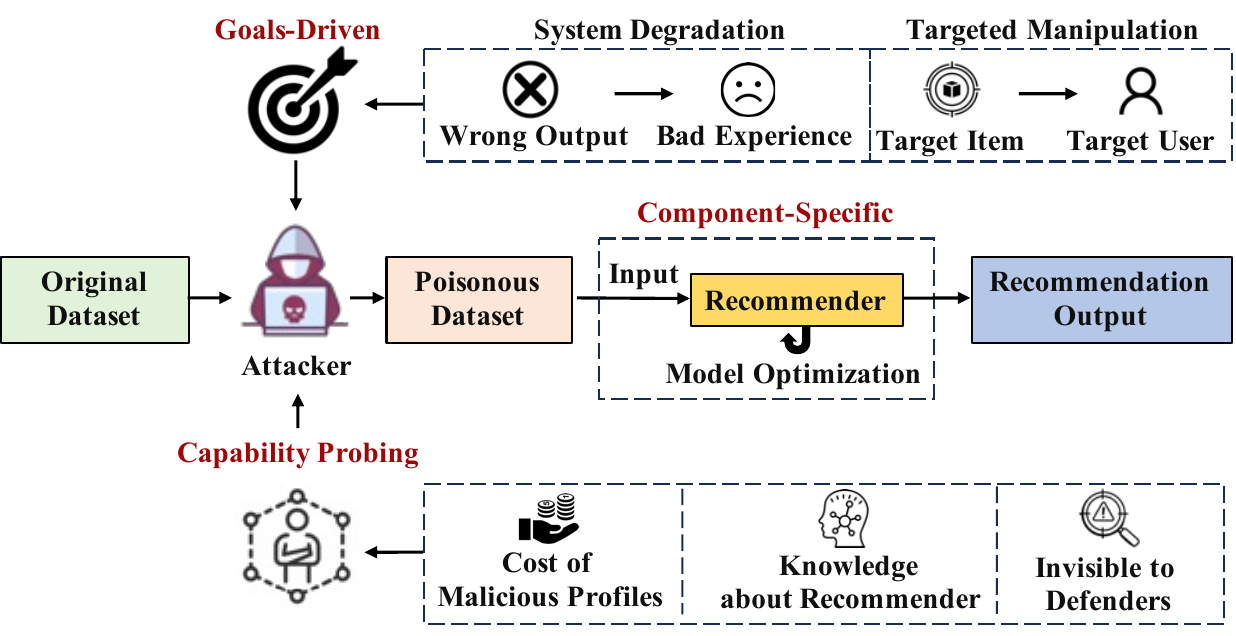}
\caption{The illustration of the proposed three types of PAR.}
\vspace{-10pt}
\label{illustration}
\end{figure}

It is important to recognize that the three identified categories of PAR within this taxonomy are interconnected rather than isolated. Specifically, the methods classified into the latter two categories frequently involve manipulating or disrupting elements within the first category as a means to their ends. Despite this overlap, each category maintains distinct characteristics, primarily distinguished by their unique focal points and objectives. This classification is critical for understanding the specific nuances and implications of each type of PAR. The illustration in Figure \ref{illustration} clearly shows the correlation of three categories, which is helpful in understanding the taxonomy. Additionally, for convenient reference, detailed tabular summaries of the surveyed literature are available online. We refer the readers to \url{https://github.com/CoderWZW/ARLib/blob/main/collection.md}.

\section{Component-Specific Attacks}
The unique characteristics of different components in RS spawn diverse forms of PAR tasks. These attacks are specifically tailored to the context of the victim system. We further divide the component-specific attacks into \textbf{input-specific}, \textbf{recommender-specific} and \textbf{optimization-specific attacks}.



\subsection{Input-Specific Attacks}
In the study of PAR, the basic input is user-item interaction data, typically represented as a tuple $(u, i, D_{ui})$, indicating that user $u$ has interacted with item $i$. This format forms the basis of most attacks in PAR. Advancing from this basic structure, graph-based RS introduce a significant shift, where users and items become nodes, connected by interactions as edges. The graph-specific PAR methods like \textit{GraphAttack} ~\cite{02fang2018poisoning} and \textit{GSPAttack} ~\cite{33nguyen2022poisoning} demonstrate vulnerabilities inherent to graph structures. They exploit item recommendation rates as their objective function, optimizing malicious user profiles by leveraging gradients derived from graph propagation dynamics. The complexity further escalates with the incorporation of heterogeneous data and knowledge graphs (KGs) ~\cite{86ji2021survey}. These elements add diverse information, such as social relationships and detailed entity attributes, like user demographics and item descriptions. \textit{MSOPDS} ~\cite{39yeh2023planning}, for example, navigates the simultaneous strategies of multiple attackers, each with unique social connections. Similarly, \textit{KGAttack} ~\cite{21chen2022knowledge} and \textit{KG-RLattack} ~\cite{36wu2022poisoning} utilize the extensive information in KGs to create synthetic user profiles that are both authentic and credible.

Temporal data, pivotal in sequential RS, also present a viable target for PAR. \textit{PoisonRec}~\cite{09song2020poisonrec} employs reinforcement learning (RL) to analyze user preferences from historical sequences, conceptualizing this process as a Markov Decision Process (MDP). This RL application facilitates the execution of PAR tasks by leveraging the sequential nature of user interactions. Due to its effectiveness, RL is extensively utilized across a range of PAR methods, such as \textit{LOKI} ~\cite{10zhang2020practical} and \textit{CopyAttack} ~\cite{08fan2021attacking}. \textit{LOKI} implements RL to train a surrogate model, facilitating the generation of user behavior samples for specific purposes, while \textit{CopyAttack} applies RL to identify and replicate real user profiles from a source domain into the target domain. This approach is strategically designed to promote a specific subset of items in the target domain, leveraging the authenticity and characteristics of real user data.

In multimodal data scenarios, vulnerabilities specific to PAR become evident. \textit{PSMU} ~\cite{29yuan2023manipulating} and \textit{IPDGI} ~\cite{83chen2023Adversarial} use guided diffusion models to create adversarial images aimed at facilitating item promotion. These models capably emulate the distribution of benign images, resulting in adversarial images that closely resemble the original ones in terms of fidelity.  Furthermore, methods like \textit{TDP-CP} \cite{80zhang2022targeted} and \textit{ARG} \cite{81chiang2023shilling} concentrate on the manipulation of textual data associated with items. They implement methods like adding, deleting, or replacing words, or even generating new textual content, with the objective of inducing shifts in recommendations. These PAR methods illustrate the extensive range of PAR across different data modalities.

\subsection{Recommender-Specific Attacks}
As RS advances, one of the research trajectories in PAR has increasingly shifted towards exploiting specific vulnerabilities within various RS architectures and paradigms. 

Starting with traditional systems, studies such as \textit{TNA} ~\cite{34fang2020influence} explore the vulnerabilities of matrix factorization-based RS, while \textit{UNAttack} ~\cite{31chen2021data} delves into neighborhood-based systems. As research advances, it extends to more complex DL-based RS. \textit{DLAttack} ~\cite{18huang2021data} focuses on pure DL-based RS, and \textit{NCFAttack} ~\cite{25zhang2020towards} addresses neural collaborative filtering-based RS. Furthermore, \textit{GOAT} ~\cite{15wu2021ready} and \textit{GSPAttack} ~\cite{33nguyen2022poisoning} execute PAR specifically on graph-based RS that predominantly incorporate graph neural networks ~\cite{60zhang2019graph} in their architectures. These studies expose vulnerabilities in the gradient update mechanisms of these DL-based systems. This evolution continues with the advent of Self-Supervised Learning (SSL). \textit{SSLAttack} ~\cite{30wang2023poisoning} targets this paradigm, shifting the focus of the PAR to the initial pre-training phase. This approach stresses the injection of adversarial influences during pre-training, which are subsequently propagated to downstream recommendation models during the fine-tuning stage. In a similar context, PromptAttack ~\cite{27wu2023attacking} explores vulnerabilities in pre-trained models utilizing prompt learning, a nuanced approach compared to traditional fine-tuning, thus exposing weaknesses in various downstream recommendation models. 

Concurrently, as concerns over privacy in centralized systems grew, research pivoted towards decentralized Federated RS, where local storage of user data enhances security and privacy. Despite this new architecture, \textit{FedRecAttack} ~\cite{19rong2022fedrecattack} demonstrates how public interactions can be exploited to approximate user features and manipulate malicious users, thus optimizing the PAR process. Building on this, \textit{A-ra} ~\cite{22rong2022poisoning} further focuses on identifying `hard users' for a target item, who are difficult to influence due to unique interaction patterns. Additionally, in response to the growing demand for transparency, explainable RS, which converts user interactions into logical expressions, has been scrutinized. \textit{H-CARS} ~\cite{28chen2023dark} leverages counterfactual explanations to craft malicious user profiles, exploiting the logical reasoning in these systems and thus uncovering new vulnerabilities.

\subsection{Optimization-Specific Attacks}
Compared to data input and recommenders, the optimization process in RS draws less attention from attackers. However, recent research found that this aspect can be particularly vulnerable to subtle yet effective manipulations.

\textit{PipAttack} ~\cite{11zhang2022pipattack} capitalizes on the distinctive characteristics of popular item aggregation within the widely used Bayesian Personalized Ranking (BPR) loss. It alters the target item to resemble the attributes of popular items in the embedding space, thereby increasing the likelihood of user interactions with that item. Additionally, \textit{CLeaR} ~\cite{82wang2023poisoning}  emphasizes that intensifying Contrastive Learning loss in RS can result in a uniform distribution of representations. This uniformity is identified as a vulnerability, amplifying the system's vulnerability to PAR.

\section{Goal-Driven Attacks}
In PAR, the nature of malicious data is highly dependent on the objectives of the attackers. As such, this category is subdivided into \textbf{system degradation attacks}, \textbf{targeted manipulation attacks}, and \textbf{hybrid-goal attacks}. 



\subsection{System Degradation Attacks}
This kind of PAR is also referred to as untargeted attacks, as they are strategically designed to disrupt the overall functionality of RS, affecting all recommendation results. The ultimate goal is to deteriorate the user experience and potentially inflict substantial financial losses on the service provider. 

To this end, one set of attack approaches centers on sampling specific users or items and then primarily poisoning the RS through these samples. For instance, \textit{Infmix} ~\cite{37wu2023influence,38wu2021fight} concentrates on identifying `influential users' among malicious users who can exert a more significant impact on the system. The approach involves prioritizing the creation of interactions for these influential users to amplify the attack's impact. Similarly, \textit{FedAttack} ~\cite{20wu2022fedattack} employs globally `hardest items' to undermine model training. It selects items most and least relevant to user embeddings as the hardest negative and positive samples, respectively. Manipulating the interaction records of these hardest items significantly influences the system's outcomes. In addition, \textit{ClusterAttack} ~\cite{23yu2023untargeted} controls poisonous profiles to cluster item embeddings into dense groups. This clustering causes the recommender to generate similar scores for items within the same cluster, thereby disrupting the ranking order. On the other hand,  \textit{UA-FedRec} ~\cite{26yi2023ua} focuses on the user and item modeling processes, adjusting the representations to increase the distance between similar samples and decrease it for dissimilar ones, thereby affecting the overall recommendation process.

\subsection{Targeted Manipulation Attacks}
Opposite to system degradation attacks, this type of PAR is often referred to as targeted attacks. These attacks are elaborately tailored to either promote or demote specific items within distinct user groups or all users.

In a typical targeted attack scenario, attackers aim to significantly influence the placement of target items in the top-K recommendation lists for as many users as possible.  To achieve this, PoisonRec ~\cite{09song2020poisonrec}, \textit{KGAttack} ~\cite{21chen2022knowledge}, and \textit{ModelExtractionAttack} ~\cite{14yue2021black} focus on maximizing the appearance frequency of targeted items in top-K lists. Alternatively, \textit{A-ra} ~\cite{22rong2022poisoning} aims to directly maximize the predicted probability of each target item being recommended. Similarly, \textit{RAPU} ~\cite{12zhang2021data} seeks to ensure the predicted probabilities of the target item exceed those of other items. In addition to directly promoting or demoting specific items, some PAR approaches emphasize unique and specialized objectives to achieve similar outcomes. A notable example is \textit{GTA} ~\cite{40wang2023revisiting}, which initially focuses on assessing user intent by predicting the most preferred item of each user and subsequently modifying the target item to closely resemble these preferred items. This operation updates the characteristics of the target item to align with the user-preferred items, subtly influencing the recommendation results. Another variant of targeted attacks in PAR is specifically designed to promote certain items to a designated user group, ensuring that these specific users are exposed to the attacker-chosen items. \textit{AutoAttack} \cite{73guo2023targeted} exemplifies this approach by considering both the target item and user perspectives. It crafts malicious user profiles that closely mimic the attributes and interactions of the targeted user group. This strategy enables an effective influence on the designated group while concurrently minimizing unintended effects on other users.

\subsection{Hybrid-Goal Attacks}
To gain more benefits, attackers might design more flexible attack strategies, aligning with multiple objectives and including both system degradation and targeted manipulation.

\textit{SGLD} ~\cite{01li2016data} stands out as a pioneering method capable of pursuing dual goals concurrently. Its core technique involves maximizing the discrepancy in all predictions before and after the poisoning attack (system degradation) and enhancing the prediction likelihood of a target item for each user (targeted manipulation). It achieves a balance between these two goals using weighted coefficients. Following the groundwork laid by \textit{SGLD}, studies such as \textit{CD-Attack} ~\cite{03chen2019data} and \textit{NCFAttack} ~\cite{25zhang2020towards} have adopted this dual-goal framework. \textit{CD-Attack} demonstrated that malicious profiles developed using this methodology exhibit good transferability across different domains, while \textit{NCFAttack} chose to customize a more elegant attack target. It is different from the former which takes the attack effect as the optimization goal. Instead, it generates specific adversarial gradients based on the model optimization strategy to enable the execution of more precise attacks.

\section{Capability Probing Attacks}
In real-world scenarios, attackers executing PAR are frequently subject to a range of constraints, which invariably influence the effectiveness of their attacks. We categorize these constraint-based PARs into three types: \textbf{knowledge-constrained probing attacks}, \textbf{cost-efficient probing attacks}, and \textbf{invisibility-assured probing attacks}. 



\subsection{Knowledge-Constrained Probing Attacks}
Assuming RS as a box, attackers' knowledge of the victim RS can be classified into three settings: white-box, gray-box, and black-box. These settings range from full access to RS, to partial system access, to scenarios where attackers only have limited interaction information and no prior knowledge of the recommendation model.

\textit{CovisitationAttack} ~\cite{04yang2017fake} theoretically evaluates three levels of attacker knowledge regarding victim models and develops different PAR strategies tailored to each level of knowledge. Furthermore, \textit{A-ra} ~\cite{22rong2022poisoning} are more focused on scenarios where attackers have limited knowledge, specifically only understanding item representation modeling while lacking knowledge of user representation modeling. To effectively implement PAR in these contexts, these approaches employ a simulated model training process to deduce user representations, adapting their strategies to the constraints of their knowledge about the victim system.

The above approaches mainly focus on the understanding of the victim model. In contrast, another line of PAR methods concentrates on the knowledge constraints pertaining to the victim dataset. \textit{AIA} ~\cite{06tang2020revisiting} and \textit{ReverseAttack} ~\cite{24zhang2021reverse} target scenarios where the attacker does not have full data access. They emphasize gathering datasets from social networking platforms to approximate user and item representations as discerned by the RS, which can then be migrated to compromise the authentic RS. \textit{RAPU} ~\cite{12zhang2021data} navigates challenges associated with data obscurities, such as data noises.  It employs a bi-level optimization framework integrated with a probabilistic generative model. This combination effectively handles data incompleteness and perturbations, allowing more accurate estimation of user and item characteristics, thereby enhancing the attack's effectiveness.

\subsection{Cost-Efficient Probing Attacks}
Each instance of manipulated or injected malicious data entails a specific cost. This expenditure is inherently constrained by factors such as the economic resources of attackers. Therefore, the optimization of PAR to achieve maximum attack effectiveness while minimizing associated costs is a critical objective for attackers.

\textit{SUI-Attack} ~\cite{72huang2023single}  addresses the constraint of a limited budget for injecting malicious users. It explores a scenario in which an attacker can only inject a single user. Despite this limitation, this approach remains effective by carefully selecting suitable items for interaction by the injected user, thereby showcasing the potential of executing PAR under tight constraints. \textit{ModelExtractionAttack} ~\cite{14yue2021black} considers scenarios with a limited budget for system output queries. It employs knowledge distillation to convert opaque, black-box victim models into transparent, white-box surrogate models that are more easily manipulable. This transformation allows the replacement of the victim models to fulfill query tasks within budgetary constraints. 

Following a similar theme of limited query budgets, \textit{CopyAttack} ~\cite{08fan2021attacking,71fan2023adversarial} and \textit{PC-Attack} ~\cite{74zeng2023practical} pre-train a surrogate model using interaction data from other multiple platforms, which aims to identify properties that are both inherent and transferable across various platforms. Subsequently, the model is fine-tuned with partial target data, customizing it for specific attack objectives. \textit{LOKI} ~\cite{10zhang2020practical} and \textit{KG-RLattack} ~\cite{36wu2022poisoning} also emphasize the use of a surrogate model, avoiding direct interactions with the target RS. The key innovation of \textit{LOKI} is the introduction of methods for directly evaluating the influence of an attack. This method efficiently assesses the impact of injected samples on recommendation outcomes without necessitating the retraining of surrogate models. In contrast, \textit{KG-RLattack} highlights the integration of rich items' attribute information from KGs into the attack strategy. It leverages publicly accessible KGs, utilizing their extensive auxiliary knowledge to significantly enhance the creation of indistinguishable malicious user profiles.

\subsection{Invisibility-Assured Probing Attacks}

To safeguard RS from PAR, most commercial RS are equipped with PAR detection methods. Consequently, executing PAR effectively necessitates careful consideration of the invisibility of malicious profiles. 

Adversarial learning is a key technology in ensuring the invisibility of malicious user profiles, achieved by aligning the distribution of adversarial fake users with that of real user interactions. \textit{AdvAttack} ~\cite{05christakopoulou2019adversarial} is the pioneering study that integrates adversarial learning into PAR. Building upon this idea, subsequent studies have utilized the more advanced framework of Generative Adversarial Networks (GANs) ~\cite{58goodfellow2020generative}, comprising generative and discriminative components. These components engage in a competitive interaction to enhance the resemblance of generated user profiles to real user profiles. Studies such as \textit{AUSH} ~\cite{07lin2020attacking}, LegUP ~\cite{32lin2022shilling}, \textit{GOAT} ~\cite{15wu2021ready} and \textit{GSPAttack} ~\cite{33nguyen2022poisoning} focus on refining the generative component to improve its efficacy. Conversely, another line of research has concentrated on enhancing the discriminative components. For instance, \textit{TrialAttack} ~\cite{13wu2021triple} incorporates an additional discriminator to assess the impact of the synthesized user profiles, offering a more detailed evaluation of the attack's influence within RS.

Beyond adversarial learning, other methods also consider profile invisibility. For example, \textit{PipAttack} ~\cite{11zhang2022pipattack} requires minimal impact on the overall recommendation performance during targeted manipulation attacks. \textit{FedRecAttack} ~\cite{19rong2022fedrecattack} imposes limits on the maximum gradient perturbations of target items post-attack, as significant perturbations could alert defenders. In a more direct approach, \textit{RecUP} ~\cite{17zhang2021attacking} and \textit{GSA-GANs} ~\cite{16wang2022gray} employ state-of-the-art malicious user detection methods to evaluate whether generated fake users can evade detection mechanisms.

\section{Challenges and Future Directions}

In this section, we shed light on the existing limitations of PAR methodologies and identify several future research directions worth exploration. This analysis aims to highlight  key areas where current PAR approaches fall short and to propose directions addressing these gaps.

\textbf{Exploration of Novel Victim Contexts.}
The emergence of innovative technologies has given rise to a multitude of recommendation scenarios, consequently expanding the landscape of potential threats. Notable advancements like multi-modal recommendation \cite{78yuan2023go},  LLM-based recommendation \cite{77wu2023survey}, and self-supervised learning-based recommendation \cite{41yu2023self} introduce unique complexities into the recommendation domain. While these advancements are rapidly advancing, research delving into these novel scenarios is still in its early phases. Each novel victim context introduces unique characteristics and vulnerabilities that deviate from those in traditional RS, revealing unexplored threats and challenges.

\textbf{Investigation of Sophisticated Malicious Intents.} Current research has identified certain attack intents, but real-world motivations for attacks are likely to be more diverse. Notably, attackers may seek to manipulate one platform's recommendations through actions on another, particularly in cases where user data are shared across platforms. Additionally, cultural attacks aim to exploit linguistic or cultural differences in multilingual RS to influence outcomes. Given the broad range of potential real-life attacks, comprehensively understanding these diverse motivations and uncovering the varied methods of PAR is critical for enhancing the security and integrity of RS.

\textbf{Theoretical Foundation for PAR.} While various PAR methods have demonstrated the vulnerability of RS, there is a lack of in-depth study on the underlying principles of PAR. For example, a key aspect is the economic balance: the quantity of malicious data injected is limited by economic factors, and similarly, the attack's effectiveness often has economic goals. This relationship between the attack's cost and its economic impact is recognized, yet its exact nature and dynamics remain unclear. The PAR domain necessitates robust theoretical foundations addressing such aspects, which would facilitate more streamlined research in PAR and reduce the dependency on laborious trial-and-error approaches.

\textbf{Long-Term Impact of PAR.} Present research in PAR tends to focus on its immediate impacts, such as the quick injection of poisoning data based on the current data and model configuration. However, this perspective may not fully capture the dynamics of RS. As RS evolve with accumulating data and ongoing model updates, the initial impact of injected poisoning data may diminish or become diluted over time. Therefore, it is essential to investigate not only the immediate effects of PAR but also to track and analyze the long-term impacts. Understanding how poisoning data interacts with and influences the RS over extended periods can provide deeper insights into the sustained vulnerabilities and resilience of these systems.

\textbf{Efficient Neutralization of PAR.} An intriguing idea under exploration is the injection of counteractive data into RS to counterbalance the negative impact of poisoning data. This strategy, while proposed in other domains ~\cite{84chan2019poison}, has been relatively unexplored in the context of PAR. The primary challenge in implementing this technique lies in achieving this neutralization effectively and efficiently, without imposing additional burdens on the system or compromising the user experience. The exploration of efficient neutralization techniques promises to be a pivotal step in enhancing the resilience of RS against sophisticated and ever-evolving poisoning strategies.

\begin{figure}[t]
\centering
\includegraphics[width=0.48\textwidth]{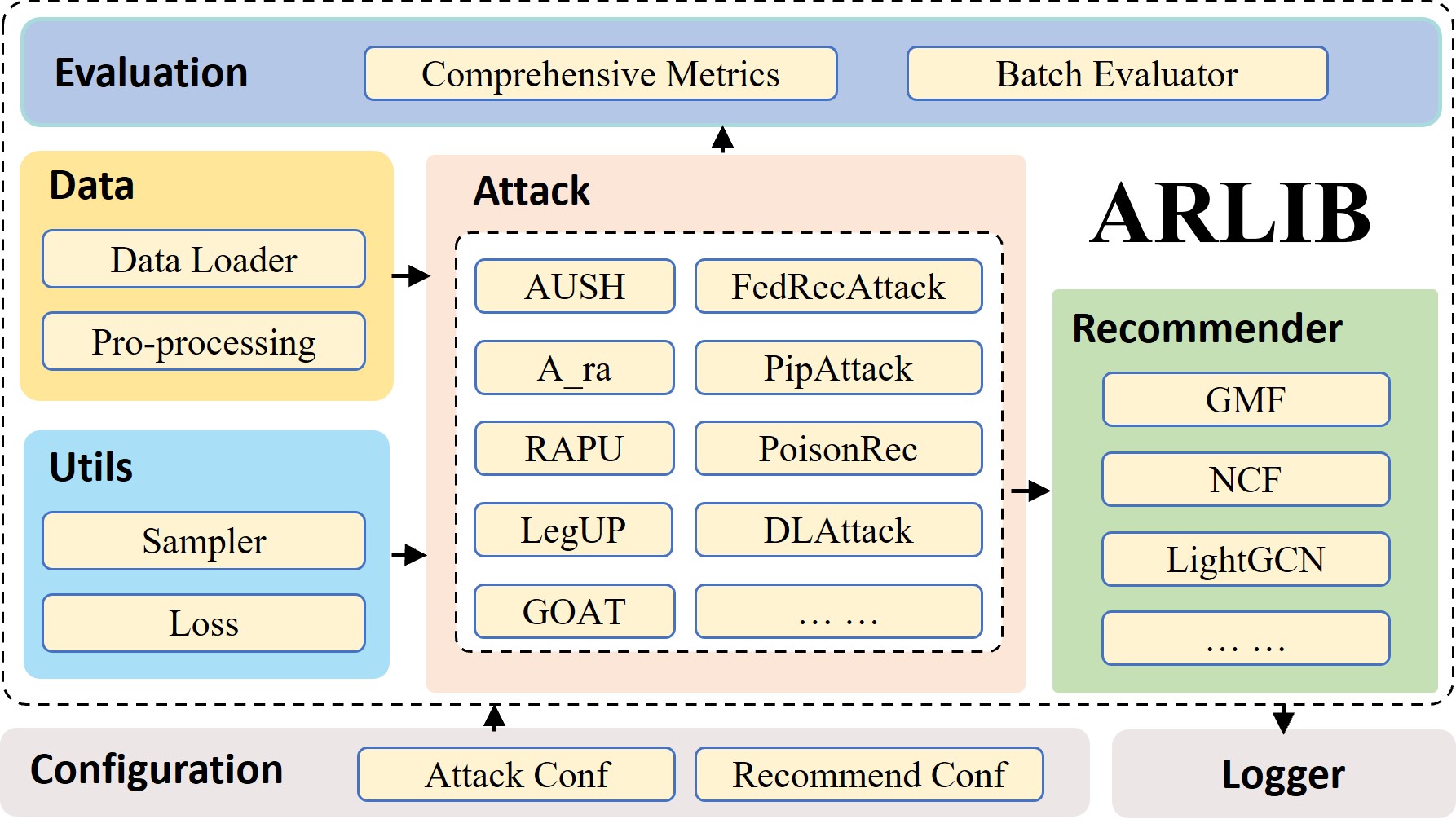}
\caption{The architecture of ARLib.}
\vspace{-10pt}
\label{framework}
\end{figure} 

\section{ARLib: A Library for Attacks against Recommendation} 
In this section, we present ARLib, an open-source library designed to facilitate the implementation of PAR. ARLib integrates over 15 different PAR models. Additionally, it incorporates all kinds of widely used datasets in PAR, catering to various research needs. While there exists related research with open-source libraries, such as ADRec ~\cite{55wang2023recad} which also offers PAR models, it primarily concentrates on detection mechanisms and includes only six PAR methods, all published before 2021. In contrast, ARLib is committed to accommodating the most recent advancements in PAR models, positioning itself as a comprehensive and continuously evolving resource for the academic community. The architecture of ARLib is shown in Figure~\ref{framework}. The code is released at \url{https://github.com/CoderWZW/ARLib}. To summarize, ARLib has the following features and functionalities:
\begin{itemize}[leftmargin=*]

\item \textbf{Fast Implementation}: ARLib is built with Python 3.8+ and PyTorch 1.8+, and is GPU-accelerated, ensuring fast development and execution. Many specialized and essential modules for PAR implementation are integrated, substantially accelerating the research pace.

\item \textbf{Modularized Architecture}: ARLib is structured into four independent core modules: Data, Recommender, Attack, and Evaluation. This modularized and unified architecture enables users to concentrate only on their method's logic rather than on irrelevant details.

\item \textbf{Highly Scalability}: Each module in ARLib is designed with user-friendly interfaces, allowing for simple to advanced customizations. This simplifies the incorporation of new recommendation methods and PAR models, endorsing a plug-and-play fashion for expansion.

\item \textbf{Comprehensive Assessment}: ARLib integrates a large number of current PAR models and provides a comprehensive suite of evaluation metrics to assess the threat of PAR models and the robustness of recommendation models against diverse poisoning attacks.

\item \textbf{Poisonous Data Simulation}: ARLib supports the customized augmentation of training datasets with artificially generated poisonous data. These enriched datasets are indispensable and the bedrock for research into PAR and robust RS.
\end{itemize}

 We have compared 15 PAR models and presented some significant findings discovered through using ARLib at \url{https://github.com/CoderWZW/ARLib/blob/main/findings.md}. These insights are believed to calibrate and guide existing and future studies on PAR, benefiting both researchers and practitioners.

\section{Conclusion}
This survey paper provides a comprehensive and detailed analysis of the latest developments in poisoning attacks against RS. It establishes a structured taxonomy that categorizes the existing methods in PAR into three distinct groups, including component-specific, goal-driven and capability probing. Furthermore, the paper outlines potential future research directions, encompassing areas such as the exploration of new victim contexts and intricate malicious intents in PAR, establishing a theoretical foundation for PAR, assessing the long-term impact of PAR, and exploring strategies for the neutralization of PAR. To facilitate practical experimentation and comparative studies, the paper introduces an open-source library offering a variety of PAR models and benchmark datasets. The availability of such a diverse range of resources is expected to significantly enhance the scope and depth of research in RS security.

\clearpage

{\small
\bibliographystyle{named}
\bibliography{ijcai23}
}

\end{document}